# AN OPEN SHOP APPROACH IN APPROXIMATING OPTIMAL DATA TRANSMISSION DURATION IN WDM NETWORKS

Timotheos Aslanidis[1] and Stavros Birmpilis[2]

[1]National Technical University of Athens, Athens, Greece
*taslan.gr@gmail.com*
[2]National Technical University of Athens, Athens, Greece
*s.birmpilis@gmail.com*

## ABSTRACT

*In the past decade Optical WDM Networks (Wavelength Division Multiplexing) are being used quite often and especially as far as broadband applications are concerned. Message packets transmitted through such networks can be interrupted using time slots in order to maximize network usage and minimize the time required for all messages to reach their destination. However, preempting a packet will result in time cost. The problem of scheduling message packets through such a network is referred to as PBS and is known to be NP-Hard. In this paper we have reduced PBS to Open Shop Scheduling and designed variations of polynomially solvable instances of Open Shop to approximate PBS. We have combined these variations and called the induced algorithm HSA (Hybridic Scheduling Algorithm). We ran experiments to establish the efficiency of HSA and found that in all datasets used it produces schedules very close to the optimal. To further establish HSA's efficiency we ran tests to compare it to SGA, another algorithm which when tested in the past has yielded excellent results.*

## 1. PROBLEM DESCRIPTION-WDM NETWORKS

In the past decade Optical WDM Networks (Wavelength Division Multiplexing) are being used quite often and especially as far as broadband applications are concerned. They provide high quality services with a large bandwidth and are therefore ideal in providing multimedia, telematic, fast internet browsing and many more communication services. In WDM network an optic fiber may be split in a number of channels depending on the frequency. Each of these channels corresponds to a specific wavelength. WDM Networks are used vastly by telecommunication companies because they increase capacity without the need to install new lines. Message packets can be interrupted using time slots in order to maximize network usage and minimize the time required for all messages to reach their destination. However, preempting a packet will result in time cost. Time to setup for the next packet can often be significant. The problem of scheduling message packets through such a network is referred to as

PBS (Preemptive Bipartite Scheduling). Even though numerous algorithms have been designed in an effort to produce efficient schedules there seems to still exist room for further research.

## 2. THE GRAPH MODEL FOR PBS

As there are 2 sets between which the multiplexing and demultiplexing process takes place the ideal representation seems to be a bipartite graph G(U,V,E,w). Source stations will be assigned to U, destination stations to V, messages to be transmitted will be the edges connecting nodes of U to nodes of V. w: E→$Q_+$ will be a weight function giving each edge e=(u,v) a weight equal to the duration of the transmission for u to v. Given a matching M in G we will denote by w(M) the maximum weight of any edge e∈M, that is w(M)=max{w(e), e∈M}. Following the notation used in previous research on the problem, Δ will denote the degree of G, W the maximum sum of edge weights incident to any of the nodes and d the setup cost to prepare for the next packet transmission. Thus, a feasible schedule for PBS would cost $\sum_{i=1}^{N} w(M_i) + d \cdot N$, where N is the number of times the network has to reconfigure so that all data will be transferred.

Using these notations, the value L=W+d·Δ represents a lower bound. L is not always achievable but is easy to calculate and is considered to be a good approximation of the optimal solution when designing near optimal algorithms for PBS.

## 3. PAST RESEARCH ON PBS

The NP-Hardness of PBS derives from the fact that it is a bicriteria minimization problem, namely the objective function to be minimized depends on two different criteria each of which affects the other. Regardless the hardness of minimizing both criteria simultaneously, minimization of each criterion separately is relatively easy. Algorithms proposed by the authors of [10] and [8] minimize the number of preemptions while the one in [15] minimizes the transmission time. In general, the problem is 4/3-ε inapproximable for all ε>0 as shown in [6]. The best guaranteed approximation ratio of any algorithm proposed for the problem is $2 - \frac{1}{d+1}$. Proof of that can be found in [1]. Many other algorithms have been proposed in order to provide solutions close to the optimal. Experimenting on test cases has yielded good results in [2], [3] and [7].

In this manuscript we try to exploit in the best way possible a reduction of PBS to the open shop scheduling problem (Om| |$C_{max}$), in order to use polynomial time algorithms proposed for some special instances of it, to minimize each criterion separately and combine the results to design a hybrid algorithm (HSA-Hybridic Scheduling Algorithm), which will tackle the bicriteria problem efficiently.

## 4. REDUCING PBS TO OPEN SHOP AND DESIGNING HSA

*Theorem 1:* Any instance of PBS can be transformed to an instance of Open Shop and vice versa.

*Proof:* Let G(U, V, E, w) be the graph corresponding a PBS instance. We transform this graph to an open shop instance in the following way: U={$u_1$, $u_2$,... $u_n$} will be the set of processors

$P=\{p_1, p_2, ..., p_n\}$. $V=\{v_1, v_2, ..., v_m\}$ will be the set of Jobs $J=\{J_1, J_2, ..., J_m\}$ and $E=\{(u_i,v_k) \mid u_i \in U, v_i \in V\}$ will be the set of operations $O=\{O_{ik} \mid i=1, 2, …, n \text{ and } k=1, 2, …, m\}$. $O_{ik}$ is the operation of $J_k$ to be processed on machine i. The processing time of each operation will be calculated by the function p: $O \to Q^+$, where $p(O_{ik}) = \begin{cases} w(u_i, v_j), & \text{if } (u_i, v_j) \in E \\ 0, & \text{otherwise} \end{cases}$.

The inverse transformation is straight forward.

Unfortunately the above reduction does not imply of a way to solve PBS using open shop algorithms as a PBS schedule would preempt all transmission simultaneously, while open shop scheduling does not have such a requirement. Yet, there exist two special instances of the Open Shop problem that are known to be solvable in polynomial time and are exactly right for our purposes. $Om|prpm|C_{max}$ in which preemption is allowed and $Om|p_{ij}=0,1|C_{max}$ in which all processing times are either 0 or 1.

The polynomial time algorithm described in [15] minimizes a preemptive open shop makespan by preempting all processor tasks simultaneously. We will refer to this algorithm by LLA (Lawler-Labetoulle Algorithm). LLA uses linear programming techniques to define a set of tasks in order to reduce the workload of all stations that, in each step of the process are assigned with the maximum workload W. The authors of [15] call this a decrementing set. The number of preemptions is $O(m^2+n^2)$. In order to improve the results of LLA instead of using a random decrementing set to reduce the workload of the stations we use one produced by a maximum weighted perfect matching algorithm. We will call this variation of LLA, POSA (Preemptive Open Shop Algorithm). We will use POSA to minimize HSA's makespan.

To complete HSA we also need an algorithm which will minimize the number of preemptions. A linear programming algorithm for $Om|p_{ij}=0,1|C_{max}$ is described in [8]. Yet, in order to better fit the requirements of our WDM network transmission we used the following Open Shop algorithm:

*OS01PT Algorithm (Open Shop 0, 1 Processing Times)*

Step1: Add the minimum number of nodes needed to G(U, V, E) so that |U|=|V|. Call the induced graph G´
Step2: Add edges to G´ to make it degree-regular.
Step3: Assign weights to the edges of G´ in the following way: Edges of the initial graph will weigh 1 while newly added edges in step2 will weigh 0.
Step4: calculate a perfect matching M in G´.
Step5: remove all edges of M from G´.
Step6: Repeat step4 and step5 until G´=∅.

To make the graph degree regular we use the subroutine described in [10].

*Theorem:* OS01PT will produce a schedule for PBS with exactly Δ transmissions.

*Proof:* By induction on the value of Δ.

For Δ=1: Since G´ is degree regular, all nodes have exactly one adjacent edge. These edges form a perfect matching for G´ and the transmission will conclude in one step.

Let the theorem stand for any regular graph with Δ=n-1.

Suppose that Δ=n. A perfect matching in G´ will reduce the degree of all nodes by one, thus making G´'s degree n-1. From the inductive hypothesis an n-1 degree graph will need n-1 transmissions to schedule its data. Therefore, to transmit all data 1+(n-1)=n transmissions will be needed. Proof that a perfect matching can always be found in a graph with Δ=n>1 can also be found in [10].

We now have all the necessary tools to design HSA:

*HSA (Hybridic Scheduling Algorithm)*

Step1: Let S1 be the feasible schedule produced for PBS using POSA. Let C1 be the cost of S1.

Step2: Let S2 be the feasible schedule produced for PBS using OS01PT. Let C2 be the cost of S2.

Step3: If C2<C1 then transmit as in S2 else transmit as in S1.

## 5. DECIDING A CRITICAL VALUE OF d FOR HSA

Five hundred test cases following a uniform distribution have been ran for a 30 source-30 destination system for values of setup cost varying from 0 to 100 and message durations varying from 0 to 120. We have to point out that since PBS is an NP-Hard problem, calculating an optimal schedule is inefficient therefore to estimate the approximation ratio we have used the lower bound to the optimal solution namely W+Δ·d.

Figure 1 depicts the deviation from the optimal solution when using POSA to calculate a schedule for PBS. Figure 2 depicts the corresponding results when using OS01PT, while Figure3 shows the results yielded by HSA. In order for OS01PT to better suit the requirements of our problem, instead of calculating a perfect matching as described in step4 of the algorithm's description, we use a maximum weighted perfect matching algorithm just as in the case of POSA.

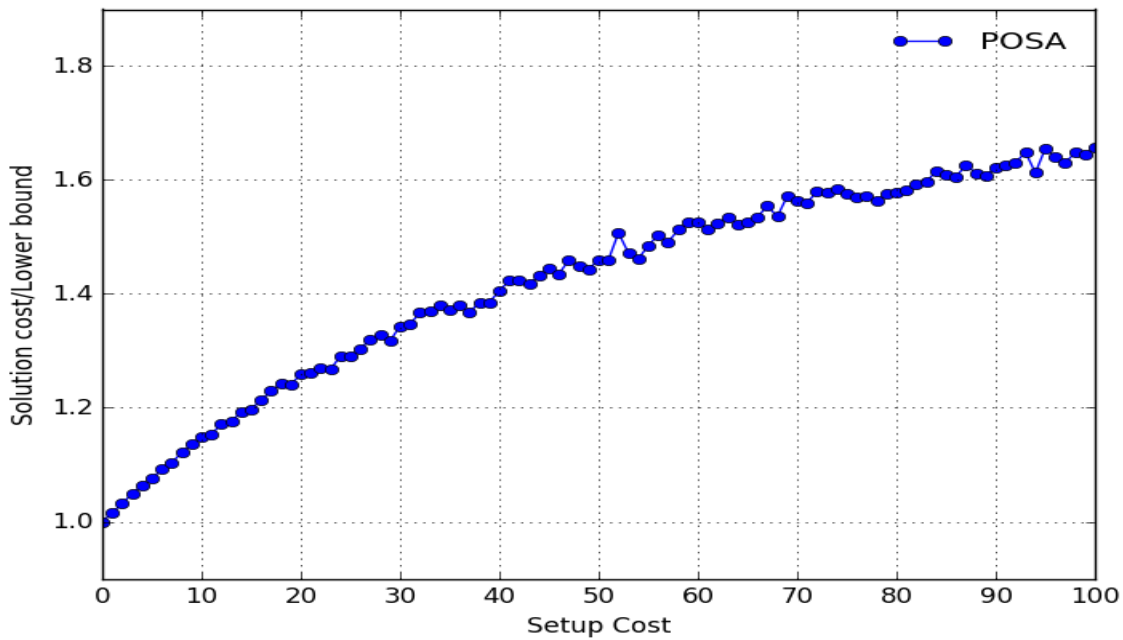

Figure 1. Average Solution cost/lower bound using POSA

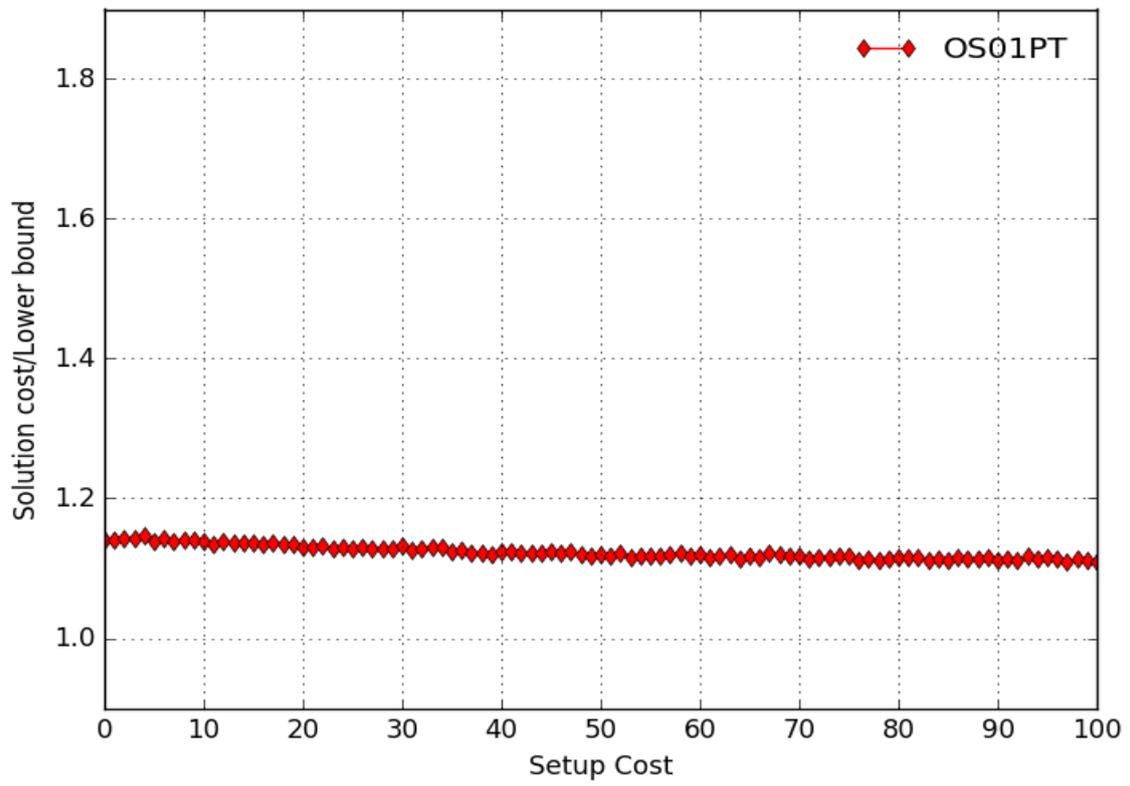

Figure 2. Average Solution cost/lower bound using OS01PT

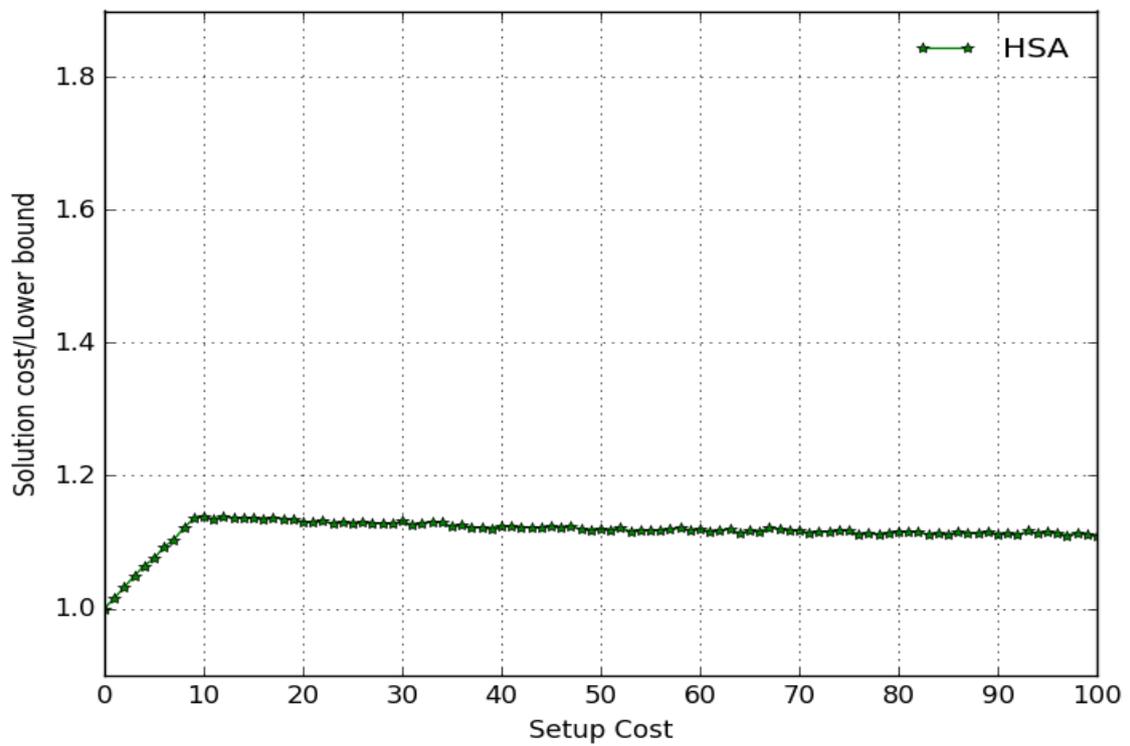

Figure 3. Average Solution cost/lower bound using HSA

According to figures1 and 2 the appropriate value of d to switch from POSA to OS01PT is d=9.

Figure 4 shows the (worst solution cost)/(lower bound) ratio of HSA for any of the instances used for each value of d. Note that it never exceeds 1.3.

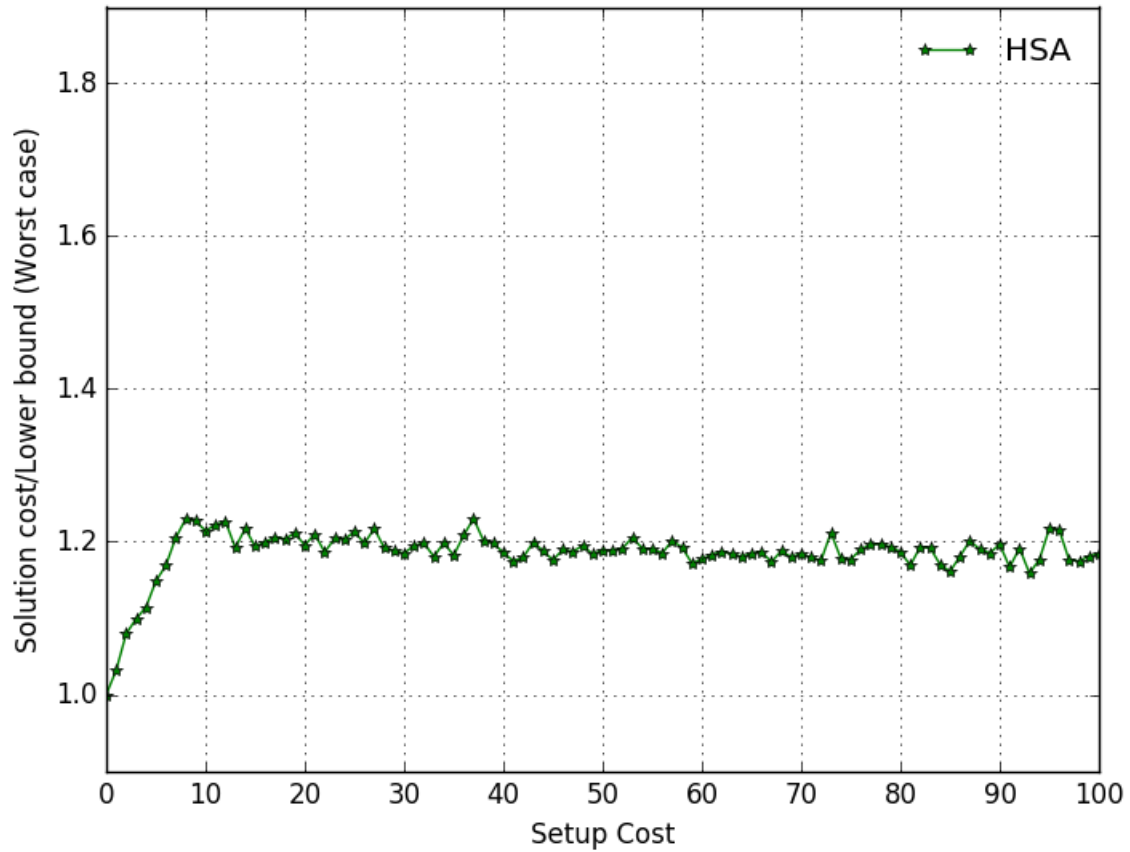

Figure 4. Worst solution cost/lower bound using HSA.

## 6. COMPARING HSA TO ANOTHER EFFICIENT ALGORITHM FOR PBS

One of the most efficient algorithms designed by researchers for PBS in the past is the SGA (Split Graph Algorithm). SGA splits the initial graph in two subgraphs, one with messages of duration less than d and another one with messages of duration at least d. The larger messages are scheduled first and then the small ones. It was found to be very efficient when tested in comparison to other efficient algorithms and it appears to be one of the top heuristics for PBS. We ran tests to compare HSA with SGA which show that HAS always produces a schedule at least as good as SGA. HSA's approximation ratio is, for some values of d up to 8% better than the one of SGA. As in paragraph 5, we used five hundred test cases following the uniform distribution for a 30 source-30 destination system for values of setup cost varying from 0 to 100 and message durations varying from 0 to 120.

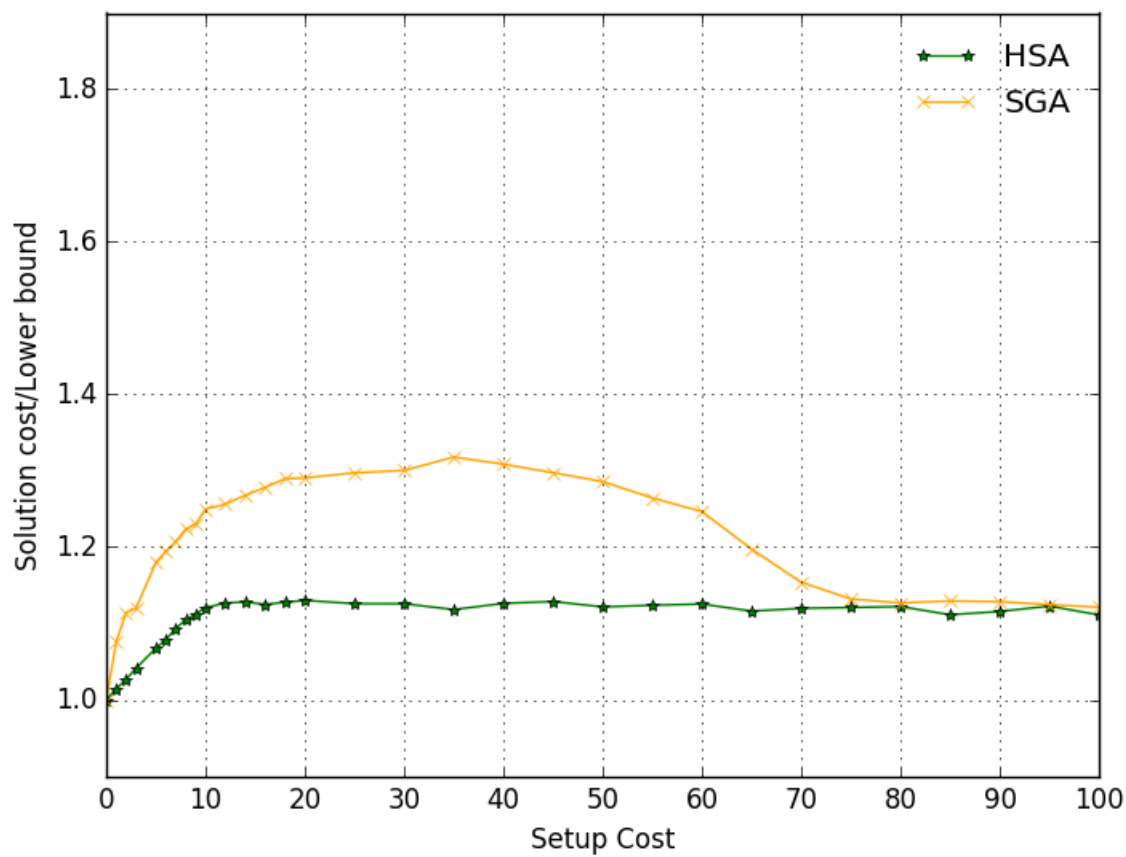

Figure 5: Comparison of HSA with SGA

## 7. CONCLUSIONS AND FUTURE WORK

In this paper we have presented a reduction of a network transmission problem (PBS) to a scheduling problem (Open Shop). Based on this reduction we have designed a hybrid algorithm (HSA) using suitable variations of polynomial time algorithms for special instances of Open Shop (POSA and OS01PT) designed for the purposes of this paper, in order to design an efficient transmission strategy for WDM network transmissions. We have ran tests to establish the efficiency of our hybrid algorithm and to suggest the appropriate value of network delay to switch from POSA to OS01PT. Knowing this value of d improves the computational complexity of HSA. In these experiments we used data following a uniform distribution. Furthermore we tested HSA against SGA, one of the most efficient algorithms designed for PBS in the past to conclude that HSA's results have in most cases an approximation ratio up to 8% better than SGA's.

Future research could focus on further improvement of the time complexity of HSA. The fact that HSA's approximation ratio even for the worst data tested has always been less than 3/2, suggests that a formal mathematical proof of an approximation ratio lower than 2 might be possible. Furthermore, tests could be ran for data following non uniform distributions such as Gaussian or Exponential. To further improve performance and complexity a new hybrid algorithm could be designed using different approaches on how to minimize each criterion separately. This algorithm might also be independent of the open shop approach. It would aim in minimizing just one of the criteria under the constraint that the other one is minimum.

**Authors**

Timotheos Aslanidis was born in Athens, Greece in 1974. He received his Mathematics degree from the University of Athens in 1997 and a master's degree in computer science in 2001. He is currently doing research at the National and Technical University of Athens in the School of Electrical and Computer Engineering. His research interests comprise but are not limited to computer theory, number theory, network algorithms and data mining algorithms.

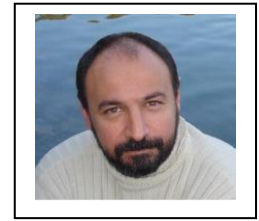

Stavros Birmpilis was born in Athens, Greece in 1994. Currently, he is a senior student in the School of Electrical and Computer Engineering at the National and Technical University of Athens, expecting to receive his diploma by July 2017. His research interests lie in the field of Algorithms and Discrete Mathematics.

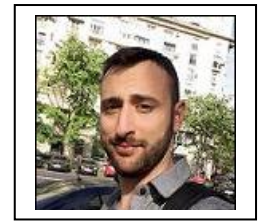